\documentclass[12pt]{article}
\usepackage{amsmath}
\usepackage{epsf}
\usepackage{epsfig}

\newcommand{\be}{\begin{equation}}
\newcommand{\ee}{\end{equation}}
\newcommand{\bear}{\begin{eqnarray}}
\newcommand{\ear}{\end{eqnarray}}
\newcommand{\ba}{\begin{array}}
\newcommand{\ea}{\end{array}}

\makeatletter

\def\compoundrel#1\over#2{\mathpalette\compoundreL{{#1}\over{#2}}}
\def\compoundreL#1#2{\compoundREL#1#2}
\def\compoundREL#1#2\over#3{\mathrel
         {\vcenter{\hbox{$\m@th\buildrel{#1#2}\over{#1#3}$}}}}
\makeatother

\begin{document}

\begin{center}

    {\Large\bf Space anisotropy search at colliders}
    \\
    \vspace{0.8cm}
    \vspace{0.3cm}
    I.~S.~Karpikov\footnote{{\bf e-mail}: karpikov@inr.ru}, 
    D.~A.~Tlisov\footnote{{\bf e-mail}: dtlisov@cern.ch},
    D.~V.~Kirpichnikov\footnote{{\bf e-mail}: kirpich@ms2.inr.ac.ru}
    \\
    
    {\small{\em 
        Institute for Nuclear Research of the Russian Academy of Sciences, }}\\
      {\small{\em
          60th October Anniversary prospect 7a, Moscow 117312, Russia
      }
      }
      \\
  \end{center}
  \begin{abstract}
In the framework of model with Lorentz violation (LV)
we discuss a physical observables for $q\bar{q}$   pair 
production at lepton-lepton colliders and describe the 
experimental signal to be detected. We obtain a 
conservative limits on Lorentz-violating dimensionless 
coupling for quark sector from LEP data. We also make a
phenomenological prediction for LV model 
at the future lepton collider. 

  \end{abstract}
\newcommand{\ve}[1]{{\bm{#1}}}
\newcommand{\vegr}[1]{{\bm{#1}}}

\section{Introduction}
The problem of space-time anisotropy is a 
great challenge of high energy physics. 
The attempts to measure the 
space anisotropy for a 
relatively low energy scale
are widely performed by an astrophysical 
experimental searches. 
Nevertheless, the ability to search fundamental 
properties of space-time on a
high energy scales appears with LHC 
launching. A violation of Lorentz 
invariance is one of the possible reason 
of space anisotropy.
There are various self-consistent setups 
of quantum gravity which admit the
violation of Lorentz invariance:
a models of quantum loop gravity 
\cite{Gambini:1998it,Alfaro:2001rb}, 
string model setups 
\cite{Kostelecky:1988zi,Kostelecky:1991ak}, a 
models of Horava-Lifshitz with extra spatial derivatives 
\cite{Horava:2009uw,Blas:2009qj,Liberati:2009pf}, 
the models of the analogue gravity~\cite{Fagnocchi:2010sn}. 
The most general Lorentz-violating Lagrangian with gauge
invariant renormalizable terms was performed by
\cite{Colladay:1998fq,Kostelecky:2003fs} for the particles
of standard model (SM). 
The former framework is known as 
Standard-Model Extension (SME) of Alan Kosteletsky. 
The current 
constraints on SME parameters 
are presented in~\cite{Kostelecky:2008ts}. 
In particular, the limits for leptons 
have been set at the level of 
$10^{-6}-10^{-20}$.


A very recent result \cite{Kostelecky:2016pyx} claims that 
$CPT$-even coefficients for LV in the quark sector 
(e.g. for $u$ and $d$ quarks) can be bounded at the level 
about $10^{-5}-10^{-6}$ from HERA experimental data on deep 
inelastic scattering (DIS) of $e^-p$. 
A framework of SME has been explored carefully also in the 
context of Tevatron collider phenomenology. In particular 
Ref.~\cite{Abazov:2012iu} provides the bounds on $CPT$-even 
LV couplings of top quark from dependence of the 
$t\bar{t}$ production cross-section on sideral time as 
the orientation of the D$0$ detector changes with the 
rotation of the Earth. Test of $CPT$-odd symmetry violation
for $B$-mesons was performed in Refs.~\cite{Schubert:2016xul,Aaij:2016mos}.

However 
the limits on $CPT$-even LV coupling $(c_{Q(U,D)})_{ZZ AB}$ 
for quarks hasn't been obtained yet.  
In the present paper we discuss a possible implication 
of the SME phenomenology for quark sector at a 
lepton- lepton colliders. Namely, we calculate the 
production 
cross-section of $q\bar{q}$ pair via 
$\gamma$ and $Z^0$-
boson for the SME couplings that affect the quark  field. 

The paper is organized as follows. In 
Sec.~\ref{LagrangianFeynmVert} we consider general 
Lagrangian of the
SME and perform LV couplings of quarks to be 
constrained by collider experiment. 
In Sec.~\ref{MatrixElemSmeSetup} we 
derive the matrix element squared for 
the process $e^+e^- \rightarrow q\bar{q}$ in SME. 
In Sec.~\ref{RefFrameTransf} we consider the spatial 
transformations from a Sun-centered reference frame to the
Earth-based laboratory frame. 
In Sec.~\ref{TimeIndepConstr} we obtain a very conservative 
limits for LV coupling of $u,d,s,c$ and $b$ quarks from ALEPH and OPAL
data. In Sec.~\ref{TimeDepConstr} we derive
time-dependent 
cross-section for the process 
$e^+e^- \rightarrow q\bar{q}$ and make 
SME prediction for the future lepton-lepton collider. 
\section{SME Lagrangian}
\label{LagrangianFeynmVert} 
We begin with a general SME Lagrangian, which can be expressed
in the following form
\be
\mathcal{L}_{SME} = \mathcal{L}_{SM} + \mathcal{L}_{LV}, 
\label{SMELagrangianGeneral}
\ee 
where $ \mathcal{L}_{SM}$ is the standard model (SM) Lagrangian
and $ \mathcal{L}_{LV}$ contains renormalizable 
Lorentz-violating terms for SM fields. Now let us consider 
CPT even Lagrangian for the quark sector
\be 
\mathcal{L}_{LV} \supset \mathcal{L}^{quarks}_{LV} = i (c_Q)_{\mu \nu A B} \overline{Q}_A \gamma^\mu 
D^\nu \overline{Q}_B + i (c_U)_{\mu \nu A B} \overline{U}_A \gamma^\mu 
D^\nu \overline{U}_B + 
\label{CTPevQuark}
\ee
$$
+ i (c_D)_{\mu \nu A B} \overline{D}_A \gamma^\mu 
D^\nu \overline{D}_B, 
$$
where index $A$ labels the quark flavor, 
$A = 1,2,3$, here $u_A = (u,c,t)$ and $d_A = (d,s,b)$.
We denote left- and right-handed quarks in 
(\ref{CTPevQuark}) by $Q_A = (u_A, d_A)_L$, 
$U_A = (u_A)_R$ and $D_A = (d_A)_R$. 
The dimensionless LV coefficients
$(c_Q)_{\mu \nu A B}$, $(c_U)_{\mu \nu A B}$
and $(c_D)_{\mu \nu A B}$ can be assumed symmetric 
in flavor indices, $A,B$ and traceless in space-time indices, 
$\mu,\nu$.
For definiteness in the present paper we consider a 
very specific 
case of Lorentz violation instead of treating full SME 
Lagrangian~(\ref{CTPevQuark}) for quarks, when $A=B$. Namely, 
the subjects of our interest are the Lagrangians for 
 quarks in the $SU(2)\times U(1)$ breaking sector,
$\mathcal{L}_{LV}=\sum_q (\mathcal{L}_{LV}^{\gamma\bar{q}q}
+\mathcal{L}_{LV}^{Z\bar{q}q})$. As an illustration
 we perform below the lagrangian for $b$ quark
\be
\mathcal{L}_{LV}^{\gamma\bar{b}b} = Q_b e \bar{b} 
\left(c_{Q\mu\nu}\frac{(1-\gamma_5)}{2}+c_{D\mu\nu}\frac{(1+\gamma_5)}{2} \right )\gamma^\mu b A^\nu 
\label{gammabb}
\ee
\be
\mathcal{L}_{LV}^{Z\bar{b}b} = \frac{e}{\sin 2 \theta_W} \bar{b} 
\left(c_{Q\mu\nu}C_L^f \frac{(1-\gamma_5)}{2}+c_{D\mu\nu}C_R^f\frac{(1+\gamma_5)}{2} \right )\gamma^\mu b Z^\nu 
\label{Zbb}
\ee
here we denote  for simplicity
$c_{Q\mu\nu} \equiv c_{Q\mu\nu 33}$ and 
$c_{D\mu\nu} \equiv c_{D\mu\nu 33}$. For other flavors only 
diagonal elements have been left, say, for  
$c$-quark we have $c_{Q\mu\nu} \equiv c_{Q\mu\nu 22}$ and 
$c_{U\mu\nu} \equiv c_{U\mu\nu 22}$.
These coefficients to be constrained by the collider 
experiment.
All remaining LV 
coefficients for 
quarks in~(\ref{CTPevQuark}) can be set to zero
without loss of generality. 
We also use 
a convenient SM notations $C_L^f = 2 T_3^f-2Q_f \sin^2 \theta_W$ 
and $C_R^f =-2Q_f \sin^2 \theta_W$ 
in (\ref{gammabb}) and (\ref{Zbb}). 


\section{The matrix element for SME setup
\label{MatrixElemSmeSetup}}
In this section we calculate the matrix element squared for 
the signal process $e^+e^- \rightarrow q\bar{q}$ at lepton-
lepton 
collider for the case of Lorentz 
violation~(\ref{gammabb}) and (\ref{Zbb}).
The amplitude squared, which 
corresponds to $q$-$\bar{q}$ pair production via $\gamma$
and $Z^0$ boson
can be written as sum of SM term and SM-SME interference terms
in the leading order of LV couplings $c^Q_{\mu\nu}$, $c^U_{\mu\nu}$ and $c^D_{\mu\nu}$ 
\be \overline{\sum_{s.c.}}|\mathcal{M}(e^+e^- \rightarrow q \bar{q})|^2 
\simeq \underbrace{ \overline{\sum_{s.c.}}|\mathcal{M}_\gamma+\mathcal{M}_Z|^2}_{ |M|^2_{SM}}+
\underbrace{ \overline{\sum_{s.c.}}\left ( 2 
\mathcal{M}_\gamma^\dagger \delta \mathcal{M}_\gamma +
4
\mathcal{M}_\gamma^\dagger \delta \mathcal{M}_Z
+
2 
\mathcal{M}_Z^\dagger \delta \mathcal{M}_Z
\right) }_{ \delta |M|^2_{SME} },
\label{MeebbTot}
\ee
in the expression above 
we average the amplitude squared over the initial state of 
lepton polarization and sum over the quark colors. 
For the sake of simplicity we now set 
$c^Q_{\mu\nu}=c^U_{\mu\nu}=c^D_{\mu \nu}\equiv c^q_{\mu\nu}$, then the partial amplitudes take the following forms 
\be
\overline{\sum_{s.c.}} 2 
\mathcal{M}_Z^\dagger \delta \mathcal{M}_Z
\!\! = \!\! \frac{2N_c e^4}{\sin^4 2\theta_W}\frac{c^q_{\mu\nu}}{(s-M_Z^2)^2}\Bigl(( C_L^{q2}\!\! + \! C_R^{q2} )(C_L^{l2}\!\! + \! C_R^{l2} )L^{\mu\nu}_V + 
( C_L^{q2}\!\! -\! C_R^{q2} )(C_L^{l2}\!\! -\! C_R^{l2} )L^{\mu\nu}_A \Bigr),
\label{ZZ}
\ee 
\begin{figure}[t]
\begin{center}
\includegraphics[width=0.5\textwidth]{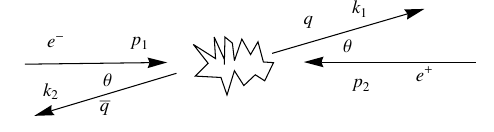}
\caption { the diagram illustrates the kinematics of the 
process $e^+e^- \rightarrow q\bar{q}$. 
\label{figure1}}
\end{center}
\end{figure} 
\be
\overline{\sum_{s.c.}} 4 
\mathcal{M}_\gamma^\dagger \delta \mathcal{M}_Z
 = \frac{2N_c e^4 Q_q Q_l }{\sin^2 2\theta_W}\frac{c^q_{\mu\nu}}{s(s-M_Z^2)}\Bigl(( C_L^q\!\! +\!\!C_R^q )(C_L^l\!\! +\!\! C_R^l )L^{\mu\nu}_V 
\!\! + \!\!( C_L^q\!\! -\!\!C_R^q )(C_L^l\!\! - \!\! C_R^l )L^{\mu\nu}_A \Bigr),
\label{GammaZ}
\ee 
\be
\overline{\sum_{s.c.}} 2 
\mathcal{M}_\gamma^\dagger \delta \mathcal{M}_\gamma
 = 2 N_c\, 2\, e^4 Q^2_q Q^2_l \frac{1}{s^2}
\, 2c^q_{\mu\nu} \, L^{\mu\nu}_V, 
\label{GammaGamma}
\ee
where $L_{\mu\nu}^V$ and $L_{\mu\nu}^A$ are the vector and 
the axial Lorentz violating tensors 
respectively, which 
depend on 
the 4-momenta of the incoming and produced particles
$e^-(p_1)e^+(p_2)\rightarrow q(k_1)\bar{q}(k_2)$: 
\be 
L_{\mu\nu}^A = ((p_2 k_1)^2-(p_2 k_2)^2) g_{\mu\nu} - (p_2 k_1)(k_{2\mu}p_{1\nu}+ k_{1\mu}p_{2\nu})+ (p_2 k_2)(k_{1\mu}p_{1\nu}+k_{2\mu}p_{2\nu}),
\label{LAmunu1} 
\ee 
\be 
L_{\mu\nu}^V =(p_2 k_1)(k_{1\nu}p_{2\mu} + k_{2\nu}p_{1\mu})\! +( p_2 k_2)(k_{1\nu}p_{1\mu} + k_{2\nu}p_{2\mu})-
\ee
$$
-(p_1 p_2)(k_{2\mu}k_{1\nu}+k_{1\mu}k_{2\nu}+p_{2\mu}p_{1\nu}+p_{1\mu}p_{2\nu})+g_{\mu\nu}(p_1p_2)^2. 
$$ 
In Sec.~\ref{TimeIndepConstr} and Sec.~\ref{TimeDepConstr}
we compare the matrix element~(\ref{MeebbTot}) to the SM 
expectation in order to estimate the contribution of 
LV coefficients $c^q_{\mu\nu}$ to the production rate of 
$q$-$\bar{q}$ pair at the lepton-lepton colliders. 

\section{The reference frame transformation
\label{RefFrameTransf}}
\begin{figure}[t]
\begin{center}
\includegraphics[width=0.48\textwidth]{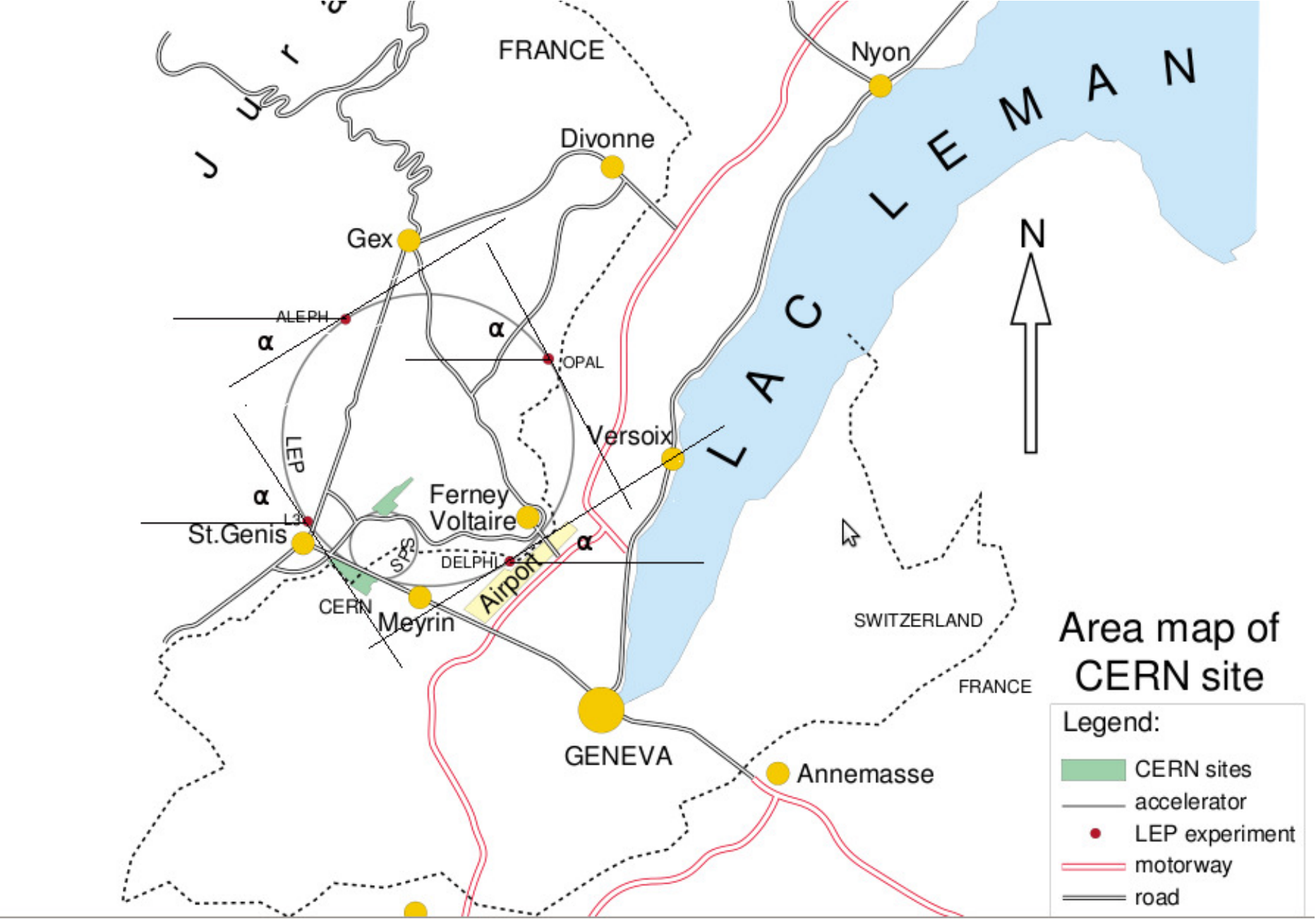}
\includegraphics[width=0.48\textwidth]{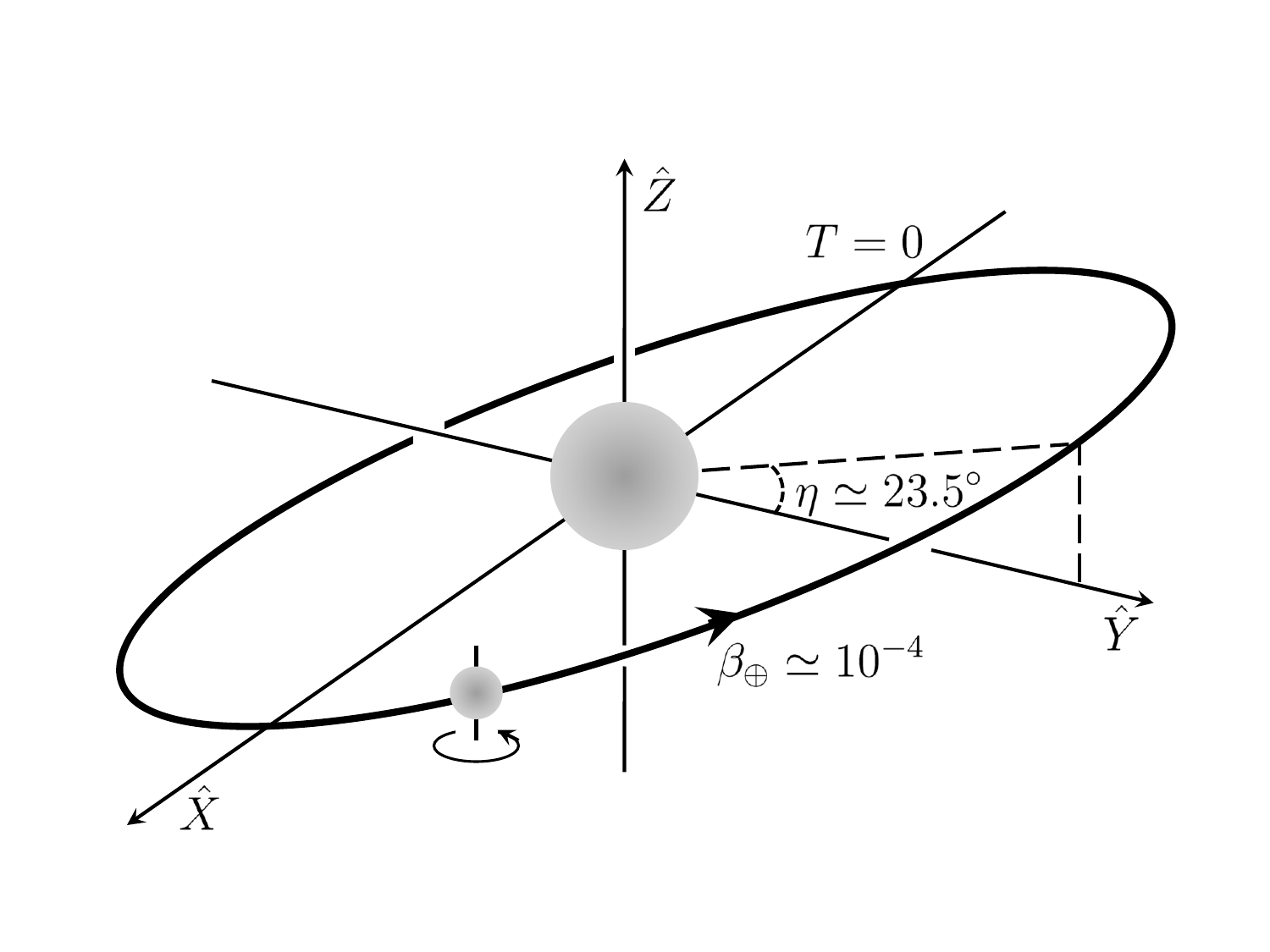}
\caption{ Left panel: orientation of the beam direction for 
ALEPH and OPAL detectors. Right panel: schematic 
illustration of the Sun-centered and Earth-based 
reference frames. 
\label{figureSun} }
\end{center}
\end{figure}

If one takes into account the Earth's rotation effect, 
then we 
should replace 
$c^q_{ij} \rightarrow c^q_{ij} (t) = c^q_{IJ}
\, R^I_i(t) R^J_j(t)$ in Eqs.~(\ref{ZZ}-\ref{GammaGamma}), the indices $I$ and $J$ 
numerate the coordinates of the Sun-centered frame,
$I,J=(X,Y,Z)$; the indices $i$ and $j$ are 
associated with 
Earth-based reference frame (see e.g.~Fig.~\ref{figureSun} 
for details). 
For the sake of simplicity we set also 
$c^q_{TT}=c^q_{TI}=c^q_{IT}=0$ throughout the paper.
 We assume that
the relative velocity
of Sun-centered and Earth-based reference frames is negligible, so the 
transformation operation involves
only rotations. The explicit form of the rotation matrix 
$\hat{R}(t)=R^J_i(t)$ is given by the following partial transformations 
\be 
\hat{R}(t) = R_z(\omega t) R_y(\chi) R_x(\pi /2) R_y(\alpha ).
\label{RTdep}
\ee 
The corresponding matrices, $ R_x(\phi)$ $ R_y(\theta)$ and 
$R_z(\psi) $ are defined by the following way
$$
R_x(\phi)= \left(\!\!
\begin{array}{ccc}
1 & 0 & 0 \\
0 & \cos \phi & -\sin \phi \\
0 & \sin \phi & \cos \phi \\
\end{array}\!\!\!
\right), 
R_y(\theta)= \left(\!\!
\begin{array}{ccc}
\cos \theta & 0 & -\sin \theta \\
0 & 1 & 0 \\
\sin \theta & 0 & \cos \theta \\
\end{array}\!\!
\right),
R_z(\psi)= \left(\!\!
\begin{array}{ccc}
\cos \psi & \sin \psi & 0 \\
-\sin \psi & \cos \psi & 0 \\
0 & 0 & 1 \\
\end{array}\!\!
\right), 
$$
here $\omega = 2 \pi/T_{sid}$ is related to the sideral period, $T_{sid}=(23\, \mbox{h}, \, 56\, \mbox{m}, \, 4.091\,
\mbox{s})$, $\chi$ is 
the colatitude of the detector, 
$\chi =(90^\circ-\mbox{Latitude})$, and 
$\alpha$ is the angle between the lepton beam and detector's 
longitude. 
\begin{table}[t]
\begin{center}
\begin{tabular}{|c|c|c|c|c|}
\hline
$ $ & ALEPH & OPAL & L3 & DELPHI \\
\hline
$ \mbox{Beam orientation} \,\, (\alpha) $ & $33.92^\circ$ & $54.50^\circ$ & $55.60^\circ$ & $34.87^\circ$ \\
\hline
$ \mbox{Colatitude} \,\, (\chi) $ & $43.77^\circ$ & $43.77^\circ$ & $43.77^\circ$ & $43.77^\circ$ \\
\hline
\end{tabular}
\end{center}
\caption{The location of the LEP detectors at Earth-based 
reference frame.
\label{BeamOrientTable}}
\end{table} 
One can see from Eqs.~(\ref{ZZ}-\ref{GammaGamma}) that
SME amplitudes squared
have the terms which are proportional to the vector part,
$c^q_{IJ}R^I_i (t)R^J_j(t)\, L^V_{ij}$, 
and to the axial part, 
$ c^q_{IJ}R^I_i (t) R^J_j (t)
\, L^A_{ij}$. Thus the effect of Earth's rotation 
will introduce a time 
dependence in the SME contribution, $\delta |\mathcal{M}|^2_{SME} (t)$, to the production rate 
of $q$-$\bar{q}$ pair. 
\section{ $q\bar{q}$ pair production at LEP
\label{TimeIndepConstr}}
The differential cross-section for $q\bar{q}$ pair production 
at LEP including Lorentz-violating contribution from SME 
can be written in the following form
\be
\frac{d \sigma }{d \Omega }(e^+e^- \rightarrow q \bar{q}) = 
\frac{1}{64 \pi^2 s} \,\, \overline{\sum_{s.c.}} \Bigl(|\mathcal{M}|^2_{SM} + \delta |\mathcal{M}|^2_{SME} (t) \Bigr).
\label{SigmaSteradian}
\ee
In this section we restrict our analysis to the case 
\be 
c^q_{IJ}= \left(\!\!
\begin{array}{ccc}
c^q_{XX}\!\! & \!\! c^q_{XY} \!\! & \!\! c^q_{XZ} \\
c^q_{YX}\!\! & \!\! c^q_{YY} \!\! & \!\! c^q_{YZ} \\
c^q_{ZX}\!\! & \!\! c^q_{ZY} \!\! & \!\! c^q_{ZZ} \\
\end{array}\!\!
\right). 
\label{CIJ1}
\ee 
The traceless condition for $c_{IJ}$ requires that
$c^q_{XX}+c^q_{YY}=-c^q_{ZZ}$. In order 
to estimate the collider sensitivity to SME coefficient we
average $\delta |M|^2_{SME}(t)$ over the sideral 
period, $T_{sid}$. 
The explicit calculation revealed that time-averaged SME 
amplitude is proportional to SM matrix element squared 
\be
\langle \delta|M|^2_{SME}(t) \rangle_t =\frac{1}{T_{sid}} 
\int\limits^{T_{sid}}_0 \, \delta|M|^2_{SME}(t) dt = C_{SME} \cdot
|M|^2_{SM} 
\label{TimeAverage}
\ee 
where 
\be 
C_{SME}= \frac{c^q_{ZZ}}{8}
(1+3(\cos 2 \alpha + \cos 2 \chi - \cos 2 \alpha \cos 2 \chi)).
\label{CSME}
\ee
Which means that SME coefficients contribute to the signal 
cross-section up to the multiplicative factor in the following 
way
\be
\sigma^{SME}_{e^+e^- \rightarrow q\bar{q}}= 
\sigma^{SM}_{e^+e^- \rightarrow q\bar{q}} \cdot (1+ C_{SME}). 
\ee
It must be point out that after time-averaging 
(\ref{TimeAverage}) our analysis is not sensitive to 
$XY$, $XZ$ or $ZY$ elements of (\ref{CIJ1}). So  we 
can constrain only $ZZ$ component of SME coupling.
Since ALEPH and OPAL detectors at LEP measured directly 
the production rate of $q\bar{q}$ events, from experimental
uncertainties on $\sigma^{SM}_{e^+e^- \rightarrow q\bar{q}}$
 we can derive 
the limits on $|c_{ZZ}^q|$ under assumption 
$|c_{ZZ}^u|=|c_{ZZ}^d|=|c_{ZZ}^s|=|c_{ZZ}^c|=|c_{ZZ}^b|=|c_{ZZ}|$, Tab.~\ref{TabLimits} shows relevant constraints.  
Beyond this assumption, namely for  $|c_{ZZ}^u|\neq|c_{ZZ}^d|\neq|c_{ZZ}^s|\neq|c_{ZZ}^c|\neq|c_{ZZ}^b|$, the 
systematic uncertainty of $b\bar{b}$~- 
and $c\bar{c}$~- pairs  fraction in total $q\bar{q}$ production
needs to be taken into account. It follows from
 $\sigma_{b\bar{b}(c\bar{c})} = R_{b(c)} \cdot \sigma_{q\bar{q}}$  that the relative uncertanties on 
$b\bar{b} (c\bar{c})$ cross-section can be expressed 
in the following way 
$\Delta \sigma_{b\bar{b}(c\bar{c})} / \langle \sigma_{b\bar{b}(c\bar{c})}\rangle = \Delta \sigma_{q\bar{q}} / \langle \sigma_{q\bar{q}}\rangle + \Delta R_{b(c)}/\langle R_{b(c)} \rangle$. 
In this case conservative bounds
can be found in Tab.~\ref{NewLimits} for ALEPH and OPAL detectors.

\begin{table}[t]
\begin{center}
\begin{tabular}{|c|c|c|}
\hline
$ $ & ALEPH    & OPAL \\
\hline
$\Delta \sigma_{q\bar{q}} / \langle \sigma_{q\bar{q}} \rangle $ & $ 0.78 \%$, see Tab.~4 of Ref.~\cite{Schael:2006wu}  &  $ 1.21 \% $, see Tab.~5 of  Ref.~\cite{Abbiendi:2003dh}  \\
\hline
$|c_{ZZ}|$  & $<0.027 $  &  $<0.036 $ \\
\hline 
\end{tabular}
\end{center}
\caption{Conservative bounds on LV coupling of 
 all quarks assuming 
 $|c^{u}_{ZZ}|=|c^{d}_{ZZ}|=|c^{s}_{ZZ}|=|c^{c}_{ZZ}|=|c^{b}_{ZZ}|=|c_{ZZ}|$.
\label{TabLimits}}
\end{table}

\begin{table}[t]
\begin{center}
\begin{tabular}{|c|c|c|}
\hline
 & ALEPH &  OPAL \\
\hline 
$\Delta \sigma_{q\bar{q}} / \langle \sigma_{q\bar{q}}\rangle$  & $0.78  \%$, see Tab.~4 of Ref.~\cite{Schael:2006wu} & $2.2 \%$,  see Tab.~2 of Ref.\cite{Abbiendi:1998ea} \\
\hline 
$\Delta R_b /\langle R_b \rangle $ & $ 9.2 \%$, see Sec.~7.1 of Ref.~\cite{Schael:2006wu} & $ 13.5 \%$, see Sec.~2.2 of Ref.~\cite{Abbiendi:1998ea} \\
\hline 
$\Delta R_c /\langle R_c \rangle $ &  $ 10.8 \%$, see Sec.~7.2 of Ref.~\cite{Schael:2006wu}   & - \\
\hline 
$|c^b_{ZZ}|$ & $<0.35$ & $<0.46 $ \\
\hline 
$|c^c_{ZZ}|$ & $ <0.4 $ &  - \\
\hline 
\end{tabular}
\end{center}
\caption{Conservative bounds on LV coupling of 
$c$- and $b$-quarks.
\label{NewLimits}}
\end{table}

\section{The prospects of SME probes in quark sector \label{TimeDepConstr}}
In this section we briefly discuss a
possible implications of the SME phenomenology for 
collider experiments and for low energy searches of LV. 
In contrast to the 
Sec.~\ref{TimeIndepConstr} now we consider the effects of 
time-dependence in the LV cross-section 
(\ref{SigmaSteradian}). In order to minimize the 
number of parameters to be constrained
we choose the following benchmark matrix of LV 
dimensionless couplings
$$
c^q_{IJ}= \left(
\begin{array}{ccc}
\!\! c^q_{XX}\!\! &\!\! 0\!\! &\!\! 0 \!\! \\
\!\! 0 \!\! & \!\! -c^q_{XX} \!\! & \!\! 0 \!\! \\
\!\! 0 \!\! & \!\! 0 \!\! & \!\! 0 \!\! \\
\end{array} 
\right). 
$$
In this case the cross-section for $q\bar{q}$ production can 
be presented as the cross-section of SM process, $\sigma^{SM}_{e^+e^- \rightarrow q\bar{q}}$, modulated 
by a time dependent function 
\be 
\sigma^{SME}_{e^+e^- \to q\bar{q}} = 
\sigma^{SM}_{e^+e^- \rightarrow q\bar{q}} \cdot (1+\epsilon(t)),
\label{sigmaTimeDep2}
\ee 
where the contribution of LV couplings is given by 
the following function
\be
\epsilon(t) = c^q_{XX} \left(\cos 2 \omega t \{\cos^2\alpha \cos^2 \chi 
-\sin^2\alpha +\cos^2\chi \}-\cos \chi \sin^2 \alpha \sin 2 \omega t \right ).
\label{epsilonT}
\ee
The expression (\ref{sigmaTimeDep2}) 
describes the variation of 
$q\bar{q}$ production signal twice with the sideral day 
due to the Lorentz violating contribution of the SME. 
We leave Monte-Carlo simulation of the 
signal~(\ref{sigmaTimeDep2}) for the future study.

 The analyses of
 Refs.~\cite{Kostelecky:2016pyx,Abazov:2012iu} are very 
sophisticated and comprehensive test of SME. However, in the 
light of prospect study, it is instructive to probe SME for 
the low energy 
observables \cite{Satunin:2017wmk} as well as 
 for the phenomenological 
quantities at the high-energy scales~\cite{Berger:2015yha}.
Indeed, Lorentz-violation in quark 
sector (\ref{gammabb}) affects the photon
polarization operator. Moreover, one can show 
that Lorentz-violating kinetic term of 
quark Lagrangian modifies the dispersion relation 
of quarks at the tree level as well as the 
photon's dispersion relation at the one-loop level 
\cite{Satunin:2017wmk}. This effectively 
means that the velocity of the photon acquires 
the additional contribution from the terms
which "run" via renormalization-group. Therefore, 
one can constrain the LV coupling of quarks with 
a high accuracy from laser experiments by measuring the 
speed of light. This sophisticated 
analysis is a subject of our study in the nearest future.
\section{Aknowledgments} 
This work was supported by RFBR grant 16-32-00803. 
We thank A.~Kostelecky, P.~Satunin, D.~Gorbunov, 
S.~Gninenko, N.~Krasnikov and V.~Matveev
 for fruitful discussions.

\end{document}